# Comparative Performance Analysis of Wireless Communication Protocols for Intelligent Sensors and Their Applications

Chakkor Saad

Department of Physics,
Team: Communication and Detection Systems,
University of Abdelmalek Essaâdi, Faculty of Sciences,
Tetouan, Morocco

Baghouri Mostafa

Department of Physics,
Team: Communication and Detection Systems,
University of Abdelmalek Essaâdi, Faculty of Sciences,
Tetouan, Morocco

El Ahmadi Cheikh

Department of Physics,
Team: Communication and Detection Systems,
University of Abdelmalek Essaâdi, Faculty of Sciences,
Tetouan, Morocco

Hajraoui Abderrahmane

Department of Physics,
Team: Communication and Detection Systems,
University of Abdelmalek Essaâdi, Faculty of Sciences,
Tetouan, Morocco

*Abstract*—The systems based on intelligent sensors are currently expanding, due to theirs functions and theirs performances of intelligence: transmitting and receiving data in real-time, computation and processing algorithms, metrology remote, diagnostics, automation and storage measurements…The radio frequency wireless communication with its multitude offers a better solution for data traffic in this kind of systems. The mains objectives of this paper is to present a solution of the problem related to the selection criteria of a better wireless communication technology face up to the constraints imposed by the intended application and the evaluation of its key features. The comparison between the different wireless technologies (Wi-Fi, Wi-Max, UWB, Bluetooth, ZigBee, ZigBeeIP, GSM/GPRS) focuses on their performance which depends on the areas of utilization. Furthermore, it shows the limits of their characteristics. Study findings can be used by the developers/ engineers to deduce the optimal mode to integrate and to operate a system that guarantees quality of communication, minimizing energy consumption, reducing the implementation cost and avoiding time constraints.

## I. INTRODUCTION

Wireless technologies have made significant progress in recent years, allowing many applications in addition to traditional voice communications and the transmission of high-speed data with sophisticated mobile devices and smart objects. In fact, they also changed the field of metrology especially the sensor networks and the smart sensors. The establishment of an intelligent sensor system requires the insertion of wireless communication which has changed the world of telecommunications. It can be used in many situations where mobility is essential and the wires are not practical.

Today, the emergence of radio frequency wireless technologies suggests that the expensive wiring can be reduced or eliminated. Various technologies have emerged providing communication differently. This difference lies in the quality of service and in some constraints related on the application and it environment. The main constraints to be overcome in choosing a wireless technology revolve around the following conditions [1], [2]:

- Range
- Reliability
- Bandwidth
- conformity (standards)
- Security
- Cost
- Energy consumption
- Speed and transmission type (synchronous, asynchronous)
- Network architecture (topology)
- Environment (noise, obstacles, weather, hypsometry)

In this work, we studies using a comparative analysis, the different parameters which influence the performance and quality of a wireless communication system based on intelligent sensors taking into our consideration the cost and the application requirements.

We can classify the requirements of applications using smart sensors into three main categories as shown in table I.

TABLE I.　　Needs Based Applications

| Types of application | Specifications and Needs |
|---|---|
| Environmental monitoring | ▪ Measurement and regular sending<br>▪ Few data<br>▪ Long battery life<br>▪ Permanent connection |
| Event detection | ▪ Alerte message<br>▪ Priority<br>▪ Confirmation statuts<br>▪ Few data<br>▪ Permanent connection |
| Tracking | ▪ Mobility<br>▪ Few data<br>▪ Localization<br>▪ Permanent connection |

## II.　Related Work

In the related work, many research studies in [3-8] have been focused on wireless sensor networks to improve communication protocols in order to solve the energy constraint, to increase the level of security and precision and to expand autonomy for accuracy, feasibility and profitability reasons. On the other side, the field of intelligent sensors remains fertile and opens its doors to research and innovation, it is a true technological challenge in so far as the topology and the infrastructure of the systems based on intelligent sensors are greatly different compared to wireless sensor networks, particularly in terms of size (number of nodes) and routing. In fact, to preserve the quality of these networks, it is very difficult even inconceivable to replace regularly the faulty nodes, which would result in a high cost of maintenance. The concept of energy efficiency appears therefore in communication protocols, [5-9]. Thus, it is very useful to search the optimization of data routing and to limit unnecessary data sending and the collisions [6], [9]. The aim challenge for intelligent sensors systems is to overcome the physical limitations in data traffic such as system noise, signal attenuation, response dynamics, power consumption, and effective conversion rates etc… This paper emphasis on the metrics of performance for wireless protocols which stands for superior measurement, more accuracy and reliability. The object of this study is for realizing an advanced intelligent sensors strategy that offers many system engineering and operational advantages which can offer cost-effective solutions for an application.

## III.　New Constraints of Intelligent Sensors System

An intelligent sensor is an electronic device for taking measurements of a physical quantity as an electrical signal, it intelligence lie in the ability to check the correct execution of a metrology algorithm, in remote configurability, in its functions relating to the safety, diagnosis, control and communication.

The intelligent sensor can be seen consisting of two parts [10-13]:

*1)　A measuring chain controlled by microcontroller*
*2)　A bidirectional communication interface with the network, providing the connection of the sensor to a central computer*

The communication part reflects all the information collected by an intelligent sensor and allows the user to configure the sensor for operation. It is therefore absolutely essential that this interface be robust and reliable. Figure 1 illustrates the intelligent sensor with its wireless communicating component. A variety of communication interfaces (wireless modules) is available, but not all sensors support these interfaces. The designer must select an interface that provides the best integration of the sensor with the others components of the system taking in our account the costs and the constraints of reliability required for a particular application.

There are others solutions to collect remote measurements such mobile and satellite communications. The main problems related to the quality of communications are: attenuation problems (distance, obstacles, rain ...), interference and multipath. The realization of the systems based on smart sensors dedicated to the applications mentioned in section I, requires the techniques and the protocols that take into account the following constraints [3]:

- The nodes are deployed in high numbers
- At any time, the nodes may be faulty or inhibited
- The topology changes very frequently
- The communication is broadcast

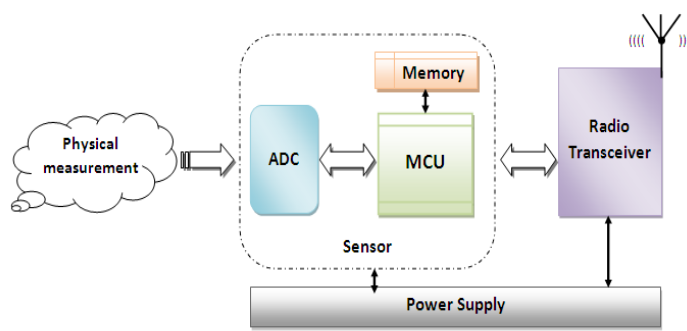

Fig. 1.　Block diagram of an intelligent sensor communication

The sensors are limited in energy, in computing capacity, and in memory. In ad-hoc networks, energy consumption was considered as an aim factor but not essential because energy resources can be replaced by the user. These networks are more focused on the QoS than the energy consumption. Contrariwise, in sensor networks, the transmission time and energy consumption are two important performance metrics since generally the sensors are deployed in inaccessible areas.

## IV.　Sensors Technology and Optimal Topology

The communication topology of the intelligent sensor systems is divided into two categories:

### A.　Direct Communication

The intelligent sensors deployed in a capture zone communicate directly with the base station via a radio link as shown in figure 2, the server collect and processes the measurements data and stores it in a database.

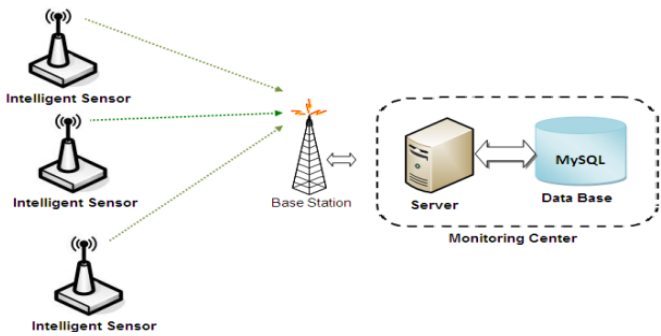

Fig. 2.   Direct communication with the monitoring center

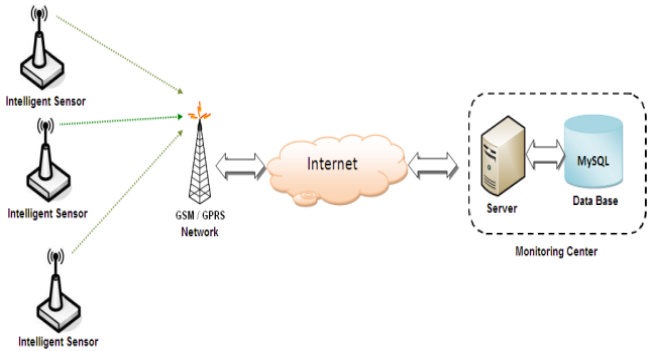

Fig. 3.   Indirect communication with the monitoring center

### B. Indirect Communication

In this case, the intelligent sensor communicates, via a GPRS network providing Internet connectivity, with the server of the monitoring center as shown in figure 3. With this architecture, it is possible to establish communications for applications that have a wider monitoring area which spreads for kilometers or when the application requires large dimensions.

### V. THE COMPARATIVE PERFORMANCE ANALYSIS

In this section, we put importance with a comparative study the following wireless protocols: Bluetooth, UWB, ZigBee, ZigBeeIP, Wi-Fi, Wi-Max, GSM/GPRS which correspond to the standards IEEE 802.15.1, 802.15.3, 802.15.4, 802.11a/b/g, 802.16 and 850-900 DCS PCS respectively [14], [15]. Based on the characteristics of each standard, obviously noticed that the UWB, Wi-Fi and Wi-Max protocols provides a higher data rate, while Bluetooth, ZigBee and GPRS provide a lower level.

Contrariwise, Bluetooth, UWB and ZigBee are intended for WPAN communication due to their range of coverage which reaches 10 m, while Wi-Fi is oriented WLAN with a range of about 100 m. However, Wi-Max and GPRS have a coverage radius of a few tens of kilometers for a WMAN network. Table II summarizes the main differences between the mentioned protocols.

TABLE II.    COMPARISON OF CHARACTERISTICS FOR DIFFERENT WIRELESS PROTOCOLS

| Protocols | Bluetooth [2], [14], [17], [18] | UWB [14], [19] | ZigBee/IP [2], [14], [17-23] | Wi-Fi [1], [2], [14], [24], [25] | Wi-Max [17], [25-28] | GSM/GPRS [29-33] |
|---|---|---|---|---|---|---|
| Frequency band | 2.4 GHz | 3.1-10.6 GHz | 868/915 MHz; 2.4 GHz | 2.4; 5 GHz | 2.4; 5.1- 66 GHz | 850/900; 1800/1900 MHz |
| Max signal rate | 720 Kb/s | 110 Mb/s | 250 Kb/s | 54 Mb/s | 35-70 Mb/s | 168 Kb/s |
| Nominal range | 10 m | 10-102 m | 10 - 1000 m | 10-100 m | 0.3-49 Km | 2-35 Km |
| Nominal TX power | 0 - 10 dBm | -41.3 dBm/MHz | -25 - 0 dBm | 15 - 20 dBm | 23 dBm | 0-39 dBm |
| Number of RF channels | 79 | (1-15) | 1/10; 16 | 14 (2.4 GHz) 64 (5 GHz) | 4;8 10;20 | 124 |
| Channel bandwidth | 1 MHz | 0.5- 7.5 GHz | 0.3/0.6 MHz; 2 MHz | 25-20 MHz | 20;10 MHz | 200 kHz |
| Modulation type | GFSK, CPFSK, 8-DPSK, π/4-DQPSK | BPSK, PPM, PAM, OOK, PWM | BPSK QPSK, O-QPSK | BPSK, QPSK, OFDM, M-QAM | QAM16/64, QPSK, BPSK, OFDM | GMSK, 8PSK |
| Spreading | FHSS | DS-UWB, MB-OFDM | DSSS | MC-DSSS, CCK, OFDM | OFDM, OFDMA | TDMA, DSSS |
| Basic cell | Piconet | Piconet | Star | BSS | Single-cell | Single-cell |
| Extension of the basic cell | Scatternet | Peer-to-Peer | Cluster tree, Mesh | ESS | PTMP, PTCM, Mesh | Cellular system |
| Max number of cell nodes | 8 | 236 | > 65000 | 2007 | 1600 | 1000 |
| Encryption | E₀ stream cipher | AES block cipher (CTR, counter mode) | AES block cipher (CTR, counter mode) | RC4 stream cipher (WEP), AES block cipher | AES-CCM cipher | GEA, MS-SGSN, MS-host |

| Authentication | Shared secret | CBC-MAC (CCM) | CBC-MAC (ext. of CCM) | WPA2 (802.11i) | EAP-SIM, EAP-AKA, EAP-TLS or X.509 | PIN; ISP; Mobility Management (GSM A3); RADIUS |
|---|---|---|---|---|---|---|
| Data protection | 16-bit CRC | 32-bit CRC | 16-bit CRC | 32-bit CRC | AES based CMAC, MD5-based HMAC, 32-bit CRC | GPRS-A5 Algorithm |
| Success metrics | Cost, convenience | Throughput, power, cost | Reliability, power, cost | Speed, Flexibility | Throughput, Speed, Range | Range, Cost, Convenience, |
| Application focus | Cable replacement | Monitoring, Data network, | Monitoring, control | Data network, Internet, Monitoring, | Internet, Monitoring, Network Service, | Internet, Monitoring, control |

## VI. Characteristics of Wireless Communication Protocols

We present in this section the different metrics to measure the performance of a wireless protocol.

### A. Network Size

The size of the GPRS network can be balanced according to the interference level, the size of data packets during traffic, the transmission protocols implemented and the number of users connected to the GSM voice services, this influences the number of GPRS open sessions which can reach 1000 to a single cell. ZigBee star network take the first rank for the maximum number of nodes that exceeds 65000, in second place there is the Wi-Fi network with a number 2007 of nodes in the BSS structure, while the Wi-Max network has a size of 1600 nodes, UWB allows connection for 236 nodes in the piconet structure, finally we found the Bluetooth which built its piconet network with 8 nodes. All these protocols have a provision for more complex network structures built from basic cells which can be used to extend the size of the network.

### B. Transmission Time

The transmission time depends on the data rate, the message size, and the distance between two nodes. The formula of transmission time in (μs) can be described as follows:

$$T_{tx} = \left( N_{data} + \left( \frac{N_{data}}{N_{maxPld}} \times N_{ovhd} \right) \right) \times T_{bit} + T_{prop} \qquad (1)$$

$N_{data}$      the data size

$N_{maxPld}$      the maximum payload size

$N_{ovhd}$      the overhead size

$T_{bit}$      the bit time

$T_{prop}$      the propagation time between two nodes to be neglected in this paper

The typical parameters of the different wireless protocols used to evaluate the time of transmission are given in Table III.

TABLE III.     Typical Parameters Of Wireless Protocols

| Protocol | Max data rate (Mbit/s) | Bit time (µs) | Max data payload (bytes) | Max overhead (bytes) | Coding efficiency+ (%) |
|---|---|---|---|---|---|
| **Bluetooth** | 0.72 | 1.39 | 339 (DH5) | 158/8 | 94.41 |
| **UWB** | 110 | 0.009 | 2044 | 42 | 97.94 |
| **ZigBee** | 0.25 | 4 | 102 | 31 | 76.52 |
| **Wi-Fi** | 54 | 0.0185 | 2312 | 58 | 97.18 |
| **Wi-Max** | 70 | 0.0143 | 2700 | 40 | 98.54 |
| **GPRS** | 0.168 | 5.95 | 1500+ | 52* | 80.86 |
| + Where the data is 10 Kbytes.     * For TCP/IP Protocol | | | | | |

From the figure 4, it is noted that the transmission time for the GSM/GPRS is longer than the others, due to its low data rate (168 Kb/s) and its long range reasons, while UWB requires less transmission time compared to the others because its important data rate.

It clearly shows that the required transmission time is proportional to the data payload size $N_{data}$ and it is not proportional to the maximum data rate.

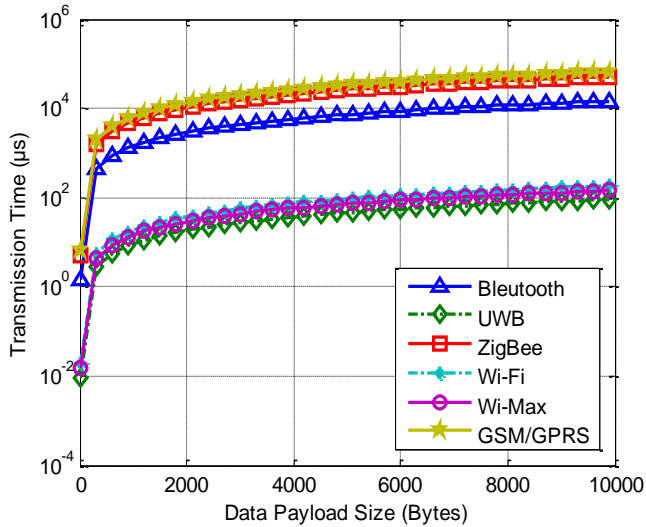

Fig. 4.   Comparison of transmission time relative to the data size

## C. Transmission power and range

In wireless transmissions, the relationship between the received power and the transmitted power is given by the Friis equation as follows [1], [33], [36-40]:

$$\frac{P_r}{P_t} = G_t G_r \left(\frac{\lambda}{4\pi D}\right)^2 \qquad (2)$$

$P_t$    the transmitted power
$P_r$    the received power
$G_t$    the transmitting omni basic antenna gain
$G_r$    the receiving antenna gain
$D$    the distance between the two antennas
$\lambda$    the wavelength of the signal

From equation (2) yields the formula the range of coverage as follows:

$$D = \frac{1}{\frac{4\pi}{\lambda}\sqrt{\frac{P_r}{P_t G_t G_r}}} \qquad (3)$$

We note that as the frequency increases, the range decreases. The figure 5 shows the variation of signal range based on the transmission frequency for a fixed power. The most revealing characteristic of this graph is the non-linearity. The signals of GSM/GPRS with 900MHz propagate much better than ZigBee, Wi-Fi, Bluetooth with 2.4GHz and UWB with 3.1GHz vice to vice coverage area.

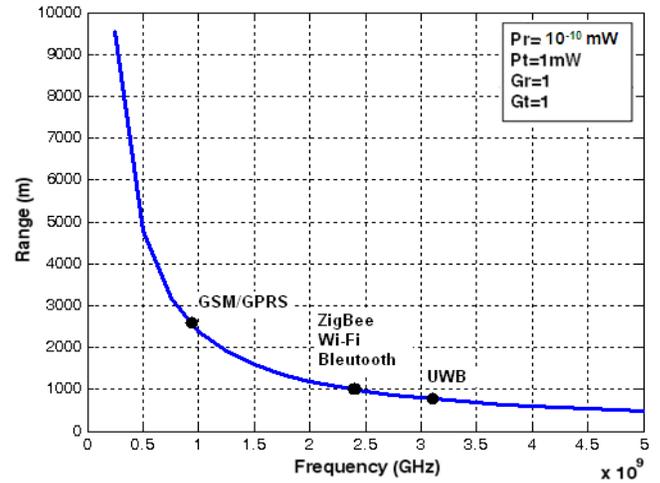

Fig. 5.   Range evolution according to the transmission frequency

## D. Energy consumption

The energy consumption for intelligent sensor involves three steps: acquisition, communication, computation and data aggregation. This consumption in the acquisition operation depends on the nature of the application [3]. Data traffic, particularly in the transmission, consumes more energy than the other operations. It also depends on the distance between the transmitter and receiver [4], [5].

The model governing the energy consumption E(p) of an intelligent sensor p depending on the communication range d(p) is given as follows:

$$E(p) = k.d^\alpha(p) + E_d \qquad \alpha \geq 2 \qquad (4)$$

$k$    the packet size
$\alpha$    the signal attenuation coefficient
$E_d$    the transmission energy costs

According to the radio energy model, [6], [38-44] the transmission power of a k bit message to a distance d is given by:

$$E_{TX}(k,d) = \begin{cases} k.\varepsilon_{fs}.d^2 + k.E_{Elec} & d < d_0 \\ k.\varepsilon_{amp}.d^4 + k.E_{Elec} & d \geq d_0 \end{cases} \qquad (5)$$

$$d_0 = \sqrt{\frac{\varepsilon_{fs}}{\varepsilon_{amp}}} \qquad (6)$$

$E_{Elec}$    electronic energy
$\varepsilon_{fs}, \varepsilon_{amp}$    amplification energy

The electronic energy $E_{Elec}$ depends on several factors such as digital coding, modulation, filtering, and signal propagation, while the amplifier energy depends on the distance to the receiver and the acceptable bit error rate. If the message size and the range of communication are fixed, then if the value of α grow, the required energy to cover a given distance increase also.

The figure 6 illustrates the evolution of the energy consumption for ZigBee protocol based on the signal range. We can say that an increase in data packet size allows then an increase of the transmission energy. The equations (4) and (5) can be generalized for the all wireless mentioned protocols. The simulation parameters are given in table IV.

TABLE IV. THE SIMULATION PARAMETERS

| Parameters | Value |
|---|---|
| $E_{Elec}$ | 50 nJ/bit |
| $\varepsilon_{fs}$ | 10 pJ/bit/m$^2$ |
| $\varepsilon_{amp}$ | 0.0013 pJ/bit/m$^4$ |

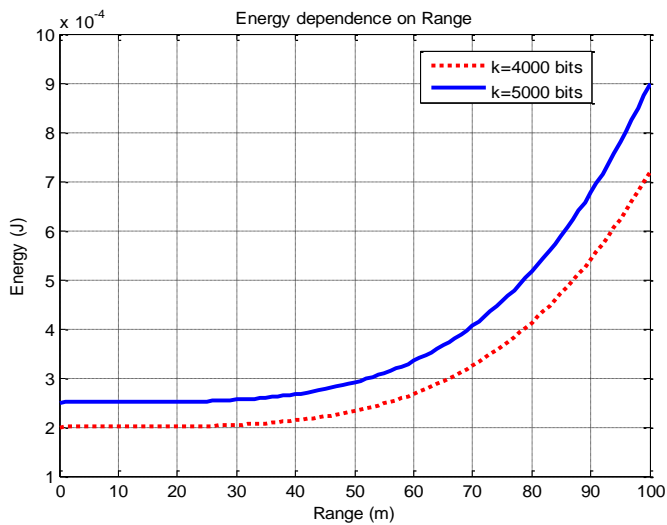

Fig. 6. The energy consumption depending on the signal range

The predicted received power by an intelligent sensor for each data packet according to the communication range d is given by the Two-Ray Ground and the Friss free space models [3], [35], [40] as follows:

$$P_r(d) = \begin{cases} \dfrac{P_t G_t G_r \lambda^2}{(4\pi d)^2 L} & d < d_c \\[2mm] \dfrac{P_t G_t G_r h_t^2 h_r^2}{d^4} & d \geq d_c \end{cases} \quad (7)$$

$$d_c = \frac{4\pi \sqrt{L} h_r h_t}{\lambda} \quad (8)$$

$L$    the path loss
$h_t$    the height of the transmitter antenna
$h_r$    the height of the receiver antenna
$d$    the distance between transmitter and receiver

The figure 7 shows the evolution of the reception power based on the signal range for the different studied protocols for a fixed data packet size:

TABLE V. THE SIMULATION PARAMETERS

| Parameters | Value |
|---|---|
| L | 1 |
| $G_t$=$G_r$ | 1 |
| $h_t$=$h_r$ | 1.5 m |

| Protocols | Transmitted Power (Watt) |
|---|---|
| Bluetooth | 0.1 |
| UWB | 0.04 |
| ZigBee | 0.0063 |
| Wi-Fi | 1 |
| Wi-Max | 0.25 |
| GSM/GPRS | 2 |

According to this figure, it is noted that when the distance between the transmitter and the receiver increases, the received power decreases, this is justified by the power loss in the path. The ZigBee, UWB and Bluetooth have low power consumption while Wi-Max, Wi-Fi and GPRS absorb more power due to theirs high communication range reason.

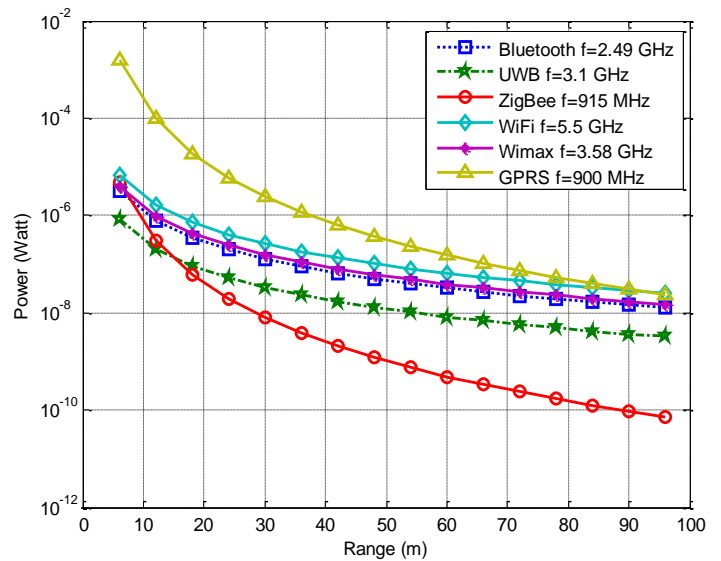

Fig. 7. The received power depending on the signal range with fixed message size

### E. Chipset power consumption

To compare practically the power consumption, we are presents in the table VI the detailed representative characteristics of particular chipset for each protocol [44-49]. The figure 8 shows the consumption power in (mW) for each protocol. Obviously we note that Bluetooth and ZigBee consume less power compared to UWB, Wi-Fi, Wi-Max and a GPRS connection. The difference between the transmission power and reception power for the protocols GPRS and Wi-Max is justified by the power loss due to the attenuation of the signal in the communication path since both of these protocols have a large coverage area.

TABLE VI.    POWER CONSUMPTION CHARACTERISTICS OF CHIPSETS

| Protocols | Chipset | $V_{DD}$ (volt) | $I_{TX}$ (mA) | $I_{RX}$ (mA) | Bit rate (Mb/s) |
|-----------|---------|-----------------|---------------|---------------|-----------------|
| Bluetooth | BlueCore2 | 1.8 | 57 | 47 | 0.72 |
| UWB | XS110 | 3.3 | ~227 | ~227 | 114 |
| ZigBee | CC2430 | 3.0 | 24.7 | 27 | 0.25 |
| Wi-Fi | CX53111 | 3.3 | 219 | 215 | 54 |
| Wi-Max | AT86 RF535A | 3.3 | 320 | 200 | 70 |
| GSM/GPRS | SIM300 | 3 | 350[*] | 230[*] | 0.164[*] |

[*] For GSM 900 DATA mode, GPRS ( 1 Rx,1 Tx )

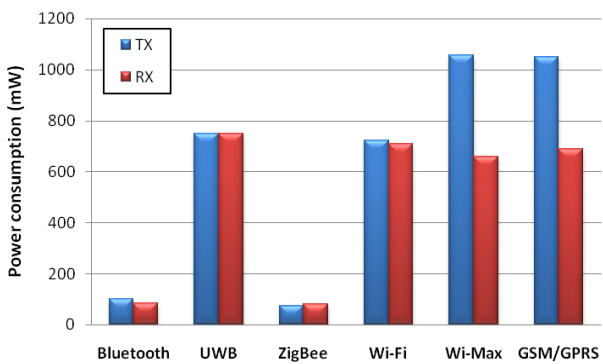

Fig. 8.    Comparison of chipset power consumption for each protocol

Based on the data rate of each protocol, the normalized energy consumption in (mJ/Mb) is shown in the figure 9, shows clearly in this figure that the UWB, Wi-Fi and Wi-Max have better energy efficiency. In summary, we can say that Bluetooth and ZigBee are suitable for low data rate applications with a limited battery power, because of their low energy consumption which promotes a long lifetime. Contrariwise for high data rate, UWB, Wi-Fi and Wi-Max would be the best solution due to their low normalized energy consumption. While for monitoring and surveillance applications with low data rate requiring large area coverage, GPRS would be an adequate solution.

*F.  Bit error rate*

The transmitted signal is corrupted by white noise AWGN (Additive White Gaussian Noise) to measure the performance of the digital transmissions (OQ-B-Q-PSK, 4PAM, 16QAM, GMSK, GFSK, 8DPSK, 8PSK and OFDM), seen in the table II, by calculating the bit error probability. The purpose of a modulation technique is not only the transfer of a data packet by a radio channel, but also achieves this operation with a better quality, energy efficiency and less bandwidth as possible.

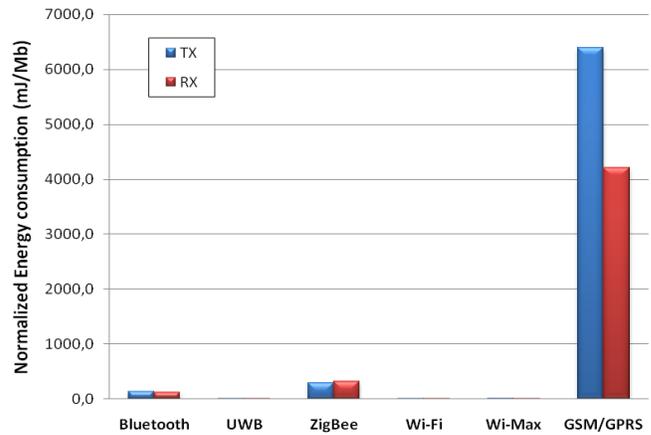

Fig. 9.    Comparing the chipset normalized energy consumption for each protocol

The bit error rate is a very good way to measure the performance of the modulation used by a communication system and therefore helps to improve its robustness. It is calculated by the following formula:

$$BER = \frac{N_{Err}}{N_{TXBits}} \qquad (9)$$

$N_{Err}$    the number of errors
$N_{TXBits}$    the number of transmitted bits

The figure 10 shows the BER of the differents modulations used in wireless technologies mentioned above according on signal to noise ratio $E_b/N_0$.

The BER for all systems decreases monotonically with increasing values of $E_b/N_0$, the curves defining a shape similar to the shape of a waterfall [36], [38]. The BER for QPSK and OQPSK is the same as for BPSK. We note that the higher order modulations exhibit higher error rates which thus leads to a compromise with the spectral efficiency.

QPSK and GMSK seem the best compromise between spectral efficiency and BER followed by other modulations. These two robust modulations are used in Wi-MAX, ZigBee, Wi-Fi and in GPRS network, can be employed in the noisy channels and in the noisy environments. However, because of their sensitivity to noise and non-linearities, the modulations 4PAM and 8DPSK remain little used compared to other modulations.

Concerning the QAM modulation, it uses more efficiently the transmitted energy when the number of bits per symbol increases; this provides a better spectral efficiency and a high bit rate. As for the frequency hopping FSK modulations, the increase of the symbols will enable reduction of the BER but also increase the spectral occupancy. The main fault of these FSK modulations is their low spectral efficiency.

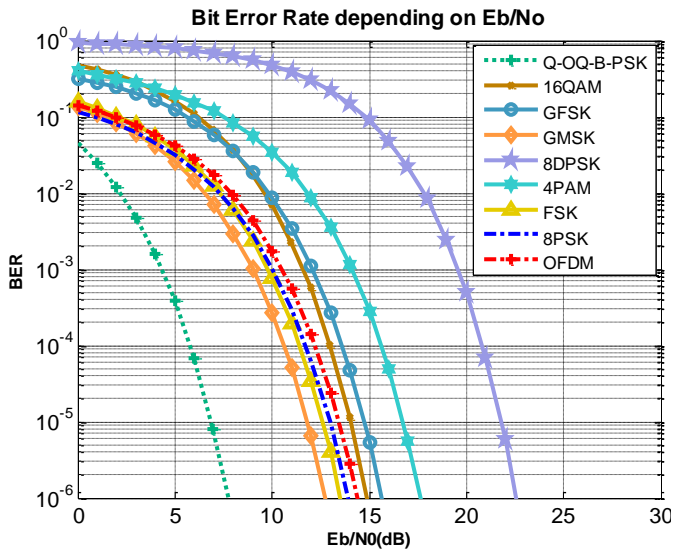

Fig. 10. Bit Error Rate for differents modulations

On the other side, the GMSK modulation has been developed in order to increase the spectral efficiency [50]. It has a satisfactory performance in terms of BER and noise resistance. This modulation is applied in the data transmission systems (MODEM), in The GSM networks [9], [35], [37], [39], [41]. The table VII gives the values of $E_b/N_0$ which cancel the BER for each modulation. Furthermore, the lower bit error probability is obtained to the detriment of the number of users. We must investigate the relationship between the transmission quality and the number of users served [50].

TABLE VII. $E_b/N_0$ VALUES WHICH CANCELS BER FOR THE DIFFERENT MODULATIONS

| Modulation | $E_b/N_0$ (dB) | B.E.R |
|---|---|---|
| B-OQ-QPSK | 7,8 | $10^{-6}$ |
| GMSK | 12,7 | $10^{-6}$ |
| FSK | 13,3 | $10^{-6}$ |
| 8PSK | 13,8 | $10^{-6}$ |
| OFDM | 14,3 | $10^{-6}$ |
| 16QAM | 14,8 | $10^{-6}$ |
| GFSK | 15,7 | $10^{-6}$ |
| 4PAM | 17,6 | $10^{-6}$ |
| 8DPSK | 22,6 | $10^{-6}$ |

### G. Data coding efficiency

The coding efficiency can be calculated from the following formula:

$$P_{cdeff} = 100 \times \frac{N_{data}}{\left(N_{data} + \left(\frac{N_{data}}{N_{maxPld}}\right) \times N_{ovhd}\right)} \quad (10)$$

Based on the figure 11, the coding efficiency increases when the data size increase. For small data size, Bluetooth and ZigBee is the best solution while for high data sizes GPRS,

UWB, Wi-Max and Wi-Fi protocols have efficiency around 94%.

In the applications point of view, for the automation industrial systems based on intelligent sensors, since most data monitoring and industrial control have generally a small size, such the pressure or the temperature measurements that don't pass 4 bytes and that don't require an important data rate, Bluetooth, ZigBee and GPRS can be a good choice due to their coding efficiency and their low data rate. On the other hand, for applications requiring a large cover zone as the borders monitoring, the persons tracking or the environmental monitoring or the event detection, GPRS and Wi-Max are an adequate solution, whereas for the multimedia applications requiring an important data rate such the video monitoring, Wi-Fi, UWB and Wi-Max form a better solution.

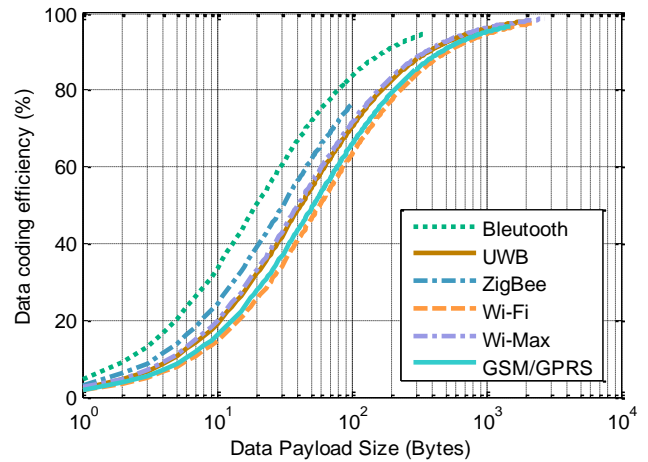

Fig. 11. Coding efficiency depending on data size

### VII. CONCLUSION

We have presented in this paper a comparative performance analysis of six wireless protocols: Bluetooth, UWB, ZigBee, Wi-Fi, Wi-Max and GSM/GPRS. However, it exists others wireless protocols as 6LoWPAN, DASH, HiperLAN…We have chosen therefore to land just the most popular ones. A quantitative evaluation of the transmission time, the data coding efficiency, the bite error rate, and the power and the energy consumption in addition of the network size permitted us to choose the best protocol which is suitable for an application based on intelligent sensor.

Furthermore, the adequacy of these protocols is influenced strongly by many others factors as the network reliability, the link capacity between several networks having different protocols, the security, the chipset price, the conformity with the application and the cost of installation that must be taking in consideration. Facing the fact that several types of wireless technologies can coexist in a capture environment, the challenge which requires is to develop a gateway (multi-standard transceiver) that enables the data exchange between these heterogeneous infrastructures with a good quality of service. This approach would allow the implementation of solutions for maintaining and for monitoring while minimizing the necessary resources and avoiding the costs associated to the compatibility testing. Solving this challenge is a perspective and a continuation of this work. It turns out that the choice of a modulation type is always determined by the constraints and the requirements of the application. The BER is a parameter which gives an excellent performance indication of a radio data link.